\documentclass[aps,pre,reprint]{revtex4-2}

\usepackage{blindtext}
\usepackage{graphicx}
\usepackage{dcolumn}
\usepackage{bm}
\usepackage{mathtools, nccmath}
\usepackage{soul}               

\usepackage{siunitx} 

\newcommand{\nM}{\,\si{\nano\textsc{m}}}

\newcommand{\uM}{\si{\micro\textsc{m}}}

\begin{document}
\title{Multispecific DNA-coatings for self-assembly}
\author{T.C.M. Stevens $^{1}$}
\author{A.v.d. Sluis $^{1}$}
\author{I.K. Voets $^{1}$}
\author{P.G. Moerman $^{1,}$} 
\email{Email: p.g.moerman@tue.nl}

\affiliation{$^{1}$ Department of Chemical Engineering and Chemistry and Institute for Complex Molecular Systems, Eindhoven University of Technology, 5600MB Eindhoven, The Netherlands}


\begin{abstract}
 DNA-coated particles are promising as building blocks for functional and finite-sized assemblies because they can be programmed with orthogonal interactions owing to the sequence-specific hybridization of DNA strands. To fully exploit this programmability, it is important to develop particles with coatings that incorporate multiple distinct DNA sequences in tunable ratios and to understand how the coating composition influences self-assembly. Here, we compared two strategies to graft multiple DNA sequences in tunable and well-defined ratios on micron-sized colloidal particles. We found that a method based on click chemistry yielded mixed coatings with large batch-to-batch variation in the composition, while a method based on isothermal DNA polymerization produced coatings of predictable composition with a precision of a few percent, but requires reaction rate measurements for each new sequence in the coating. Our self-assembly experiments showed that, even with precise control over coating composition, equilibrium co-assembly of multiple types of DNA-coated particles is limited by the number of interactions that are reversible within the same narrow temperature window. This finding highlights the need to explicitly incorporate sequential assembly pathways into structure design, with coating composition dictating the order of binding events, Together, our results show how systematic tuning of interaction strength and sequential assembly through multispecific DNA coatings is a prerequisite for the experimental realization of finite-sized and dynamic structures that have so far remained largely theoretical.   
\end{abstract}

\maketitle

\section{Introduction}

DNA-coated colloids are versatile building blocks for the programmable self-assembly on the nano- and micrometer scale with the ability to form complex, responsive, and reconfigurable structures \cite{jacobs_assembly_2025, zhang_programming_2021}.  Underlying this programmability is the sequence-specific hybridization of short DNA strands that facilitates selective attractions between particles coated with complementary sequences \cite{kahn_designer_2022, jacobs_assembly_2025, laramy_crystal_2019}. Combining the sequence specificity of DNA with other design parameters such as directional interactions \cite{feng_dna_2013,oh_reconfigurable_2020, kahn_designer_2022}, particle shape \cite{jones_dna-nanoparticle_2010}, surface mobility of the DNA \cite{rinaldin_colloid_2019}, and kinetic control over the assembly process, researchers have assembled a range of structures, such as crystals \cite{he_colloidal_2020, wang_crystallization_2015}, quasicrystals \cite{zhou_colloidal_2024}, metamaterials \cite{shelke_flexible_2023}, gels, and lower dimensional structures, such as chains, rings, and stick figures \cite{leunissen_switchable_2009,valignat_reversible_2005,liu_self-organized_2016}. These studies show that DNA coatings can be used to organize particles with varied chemical make-up into precisely micro-structured materials\cite{jacobs_assembly_2025,tian_prescribed_2015}. 

The DNA coating on a single microparticle contains tens of thousands of strands \cite{oh_high-density_2020, wang_crystallization_2015, kim_engineering_2006, zheng_hopping_2024}, and in principle has the information capacity to encode on the order of 100 orthogonal interactions with potential binding partners after accounting for geometric, entropic, and sequence overlap effects\cite{wu_polygamous_2012}.  The ability to encode such a large number of selective and orthogonal interactions on a single building block, and the large self-assembly design space it offers, is the central promise of DNA-mediated assembly \cite{rogers_using_2016, jones_programmable_2015}. Indeed, simulations showed that mixtures of such multi-specific (also called promiscuous, or polygamous\cite{wu_polygamous_2012}) particles, can assemble into complex and finite-sizes structures\cite{zeravcic_size_2014} and are key to making clusters with dynamic properties such as shape shifting, catalysis, and self-replication \cite{mahynski_grand_2020,zeravcic_self-replicating_2014}. 

While experimental studies demonstrated the possibility of encoding multiple interactions on a single particle \cite{wu_polygamous_2012, di_michele_multistep_2013, oh_reconfigurable_2020, seyforth_underappreciated_2025, leunissen_switchable_2009} and opened the door to multifarious self-assembly \cite{murugan_multifarious_2015}, a gap remains between the number of distinct binding partners successfully implemented in experiments and those envisioned in simulations. Experiments showed that even with few sequences per particle, the competition between intra- and inter-particle DNA bond formation can result in interesting behaviors, such as binding only after prolonged contact \cite{leunissen_switchable_2009, sakamoto_assembly_2017}, re-entrant melting \cite{angioletti-uberti_re-entrant_2012}, or reconfiguration of crystal structures \cite{oh_reconfigurable_2020}. Simulations on the other hand showed that increasing the number of orthogonal interactions per particle makes it possible to self-assemble finite-sized, aperiodic architectures \cite{zeravcic_size_2014, mahynski_grand_2020,zeravcic_self-replicating_2014}.

Orthogonal interactions alone are not sufficient to realize the structural complexity seen in simulations. Equilibrium assembly of DNA-coated colloids requires that the interaction strengths of all binding partners be tuned such that every interaction is reversible at the same temperature. A difficult task because the temperature window for reversible interactions between DNA-coated colloids is notoriously narrow -- on the order of 1 °C \cite{wang_crystallization_2015, cui_comprehensive_2022, wu_polygamous_2012, jo_dna-coated_2024} -- and depends not only on the individual binding energy of each sequence but also on the number of strands of each sequence in the DNA brush. For successful equilibrium assembly of multispecific particles, these temperatures must be synchronized across all sequences involved, a condition that demands strategies for fabricating colloids with mixed DNA coatings whose compositions can be precisely controlled.

Here, we show how DNA sequence composition can be tuned on multispecific colloids and how closely the resulting coatings match their intended compositions. We compared two strategies: (i) simultaneous grafting of a mixture of functionalized DNA sequences, whose composition is often assumed to reflect the input ratio \cite{wang_crystallization_2015, zheng_hopping_2024, oh_reconfigurable_2020, wu_polygamous_2012, di_michele_multistep_2013, leunissen_switchable_2009, maye_dnaregulated_2007, dreyfus_aggregation-disaggregation_2010}, but which is rarely verified \cite{zheng_hopping_2024, dreyfus_aggregation-disaggregation_2010}, and (ii) appending new binding domains onto pre-grafted strands by isothermal DNA polymerization  \cite{moerman_simple_2023, maye_dnaregulated_2007}. We then determined how coating composition influences assembly pathways and structures. Together, these results reveal under what conditions multispecific particles assemble simultaneously and reversibly, and when sequentially and by diffusion-limited aggregation, thereby establishing key parameters that govern the self-assembly of complex DNA-programmed colloidal materials.

\section{Results}

\subsection{Coating Composition Control with DBCO-Click Chemistry}

\begin{figure*}[ht]
 \centering
 \includegraphics[width=17.1 cm,trim=0 420 0 0,clip]{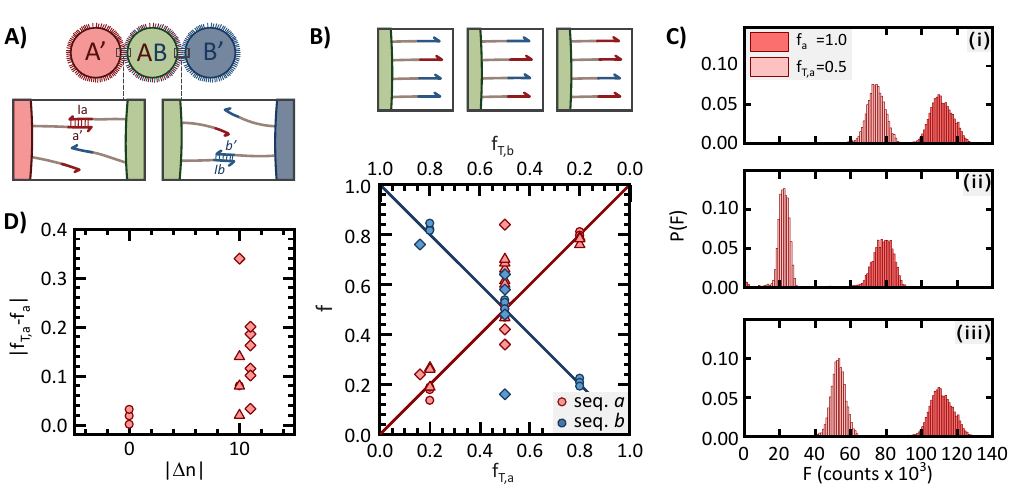}
 \caption{ (A) Schematic representation of multispecific interactions. Particles of type \textit{AB}, functionalized with two distinct DNA sequences, can simultaneously bind to complementary \textit{A}' and \textit{B}' particles. On these multispecific particles, the relative grafting densities of the two DNA sequences can be tuned. (B) Dual grafting of distinct DNA sequences onto single particles via DBCO–click chemistry was achieved by mixing different ratios of DBCO-functionalized DNA strands. The graph compares the target (input mixing) ratios with the measured grafting fractions for three different sequence combinations. In all cases, sequence \textit{a} (red) was held constant and paired with (i) a poly-thymine sequence lacking a sticky end (triangles), (ii) a sequence \textit{b} of the same length (circles), or (iii) an extended version of \textit{b} (rhombus). The data show good agreement between target and measured grafting at low and high target ratios, with increased variability observed near equal grafting fractions. (C) Fluorescence histograms of 10,000 individual particles as measured for sequence $a$ for each of the sequence combinations of panel B, all targeting $f_{T,a} = 0.5$, compared to reference particles grafted at $f_a = 1$. Intensity corresponds to grafting of sequence $a$. (D) Deviation from $f_{T,a}=0.5$ versus the nucleotide length difference $\Delta n$ between sequence $a$ and its partner.
 }
 \label{fig: DBCO}
\end{figure*}

A straightforward way to make multispecific DNA-coated colloids (i.e. particles with multiple sequences in their coating, Fig. \ref{fig: DBCO}A) is to mix functionalized DNA strands and graft the mixture onto particles. If the grafting reaction is independent of sequence and local chemical environment, the composition of the mixture should directly translate to the composition of the coating. This strategy is not only used to produce multispecific particles \cite{oh_reconfigurable_2020-1, wu_polygamous_2012, di_michele_multistep_2013, leunissen_switchable_2009,maye_dnaregulated_2007}, but also to tune the grafting density of a single sequence by mixing it with an inert strand such as poly-T \cite{wang_crystallization_2015, zheng_hopping_2024, wu_polygamous_2012, di_michele_multistep_2013, dreyfus_aggregation-disaggregation_2010}. Depending on the surface chemistry, the DNA may be biotin-labeled and react with streptavidin-coated surfaces, or DBCO-labeled and react with azide-functionalized particles, but in either case it is unclear whether the assumption that the DNA mixture determines the DNA coating composition holds. Here, we focus on DBCO-labeled DNA, which is known to yield dense, homogeneous coatings suitable for self-assembly \cite{oh_high-density_2020}.

 To measure how well the composition of the DNA coating on the particles' surfaces reflects the mixing ratio of DNA strands in the reaction mixture, we grafted particles using the mixed-DNA click method for a single functional sequence, \textit{a}, mixed with an inert poly-T strand, quantified the fraction of surface grafted DNA with sequence \textit{a}, $f_a$, and compared the obtained ratio with the target ratio. We targeted grafting fractions, $f_{T,a}$, of 0.2, 0.5 and 0.8, using that the total DNA density on the surface of the particle $\rho_{max} \approx 10^5$ strands/particle (SI \ref{SI: DBCO/seq dependent grafting density}). In order to quantify the grafting density, we hybridized the particles with complementary DNA strands with a fluorescent group and measured the fluorescent signal per particle using flow cytometry (see Methods). 

 Fig. \ref{fig: DBCO}B shows the measured fractions as a function of the targets for three separate syntheses (triangles). We found that, while the average measured fraction matches the target, there is large batch-to-batch variation, particularly at a target ratio of 0.5, indicating that grafting a mixture of two DNA strands in a 1:1 ratio onto the particle surface is not guaranteed to produce particles with the same DNA ratio in their DNA brush. 

To investigate the origin of the batch-to-batch variation we first looked at how the composition of the DNA brushes varies over the particle population in one sample using the particle's fluorescence intensity as a proxy for the number of strands in their coating. We found that the distribution of fluorescence values was comparable in width for reference particles with pure DNA coatings and for mixed-coating particles (Fig. \ref{fig: DBCO}Ci). This finding indicates that the large spread in Fig. \ref{fig: DBCO}B stems from variation between samples, not between particles within a sample.

Next, we asked whether the length of the DNA strands affects the coating composition and found that the degree of batch-to-batch variation depended on the length difference between the two DNA strands (Fig. \ref{fig: DBCO}B,D). The spread around the target composition was large when mixing a strand $a$ with a shorter polyT sequence (triangles), but mixing two strands of comparable length, binding energy, and nucleotide composition produced coating compositions that closely matched the expected trend (circles). Mixing sequence \textit{a} with an 11-nucleotide longer DNA sequence $b$ (rhombus) again resulted in variable grafting fractions, similar to what we found for mixing sequence \textit{a} with an inert poly-T sequence. Surprisingly, a length mismatch did not cause one of the strands to be systematically overrepresented on the particle surface, which would be expected if the adsorption efficiency or reactivity of the strands depended on their length -- such dependence was reported for similar click reactions on silica nanoparticles \cite{siegel_universal_2024}, and we indeed found that the overall grafting density depends on sequence (SI \ref{SI: DBCO/seq dependent grafting density}). The variation in coating compositions within a sample was also independent of the length of the DNA strands (Fig. \ref{fig: DBCO}C); suggesting that batch-to-batch variation is not a reflection of particle-to-particle variation.  These findings suggest that the spread around the target coating composition comes from a stochastic effect on the scale of the sample that somehow is more prevalent for strands of different sizes than for strands of the same size, but we do not understand by what mechanism.

Together, our results show that the mixing ratio of DNA strands in the solution cannot be taken as a proxy for the DNA in the coating on the particle, and present a need for alternatives to the mixed-DNA click method to engineer multicomponent DNA coatings with precisely tunable mixing ratios for fully addressable self-assembly.

\subsection{Coating composition control with the primer-exchange reaction}

\begin{figure}
 \centering
 \includegraphics[width=8.3 cm,trim=0 220 0 0,clip]{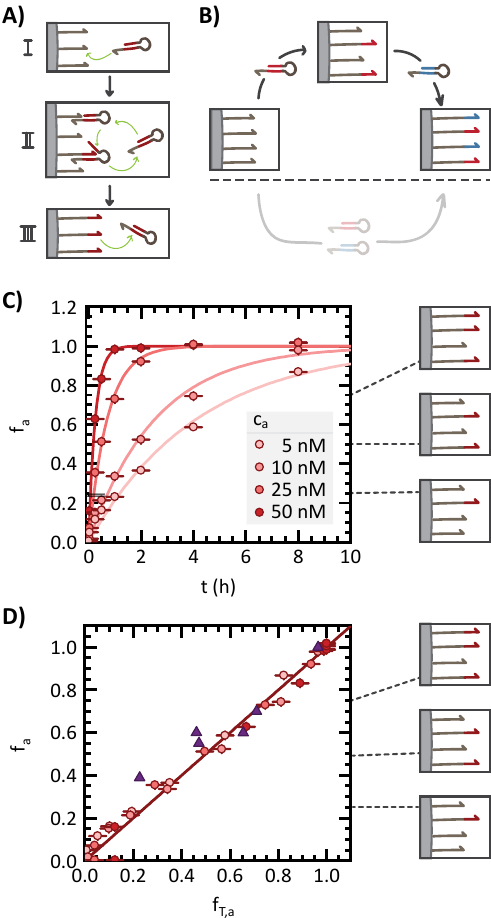}
 \caption{(A) Primer exchange reaction (PER) mechanism. (I) A DNA template hybridizes to the particle‑grafted strand. (II) The strand is extended by template‑directed isothermal DNA polymerization. The template’s hairpin enables reversible binding, allowing reuse on the next strand. (III) The reaction ends once all input strands are extended. (B) Multispecific particles produced by multi‑step PER. In step one, a sequence is grown and the reaction stopped early, leaving unreacted sites. In step two, these sites are extended with a different sequence via template exchange. (C) The grafted fraction of new DNA domains grows exponentially over time, increasing faster at higher template concentrations. (D) Data from panel C replotted against the expected fractional conversion (solid line). Expected fractions were calculated from the reaction time, template concentration, and the mean reaction rate from three measurements. Purple triangles show independent reaction batches, distinct from panel C, run to reach specific target fractions.}
 \label{fig: PER}
\end{figure}

For more precise control over the composition of mixed DNA coatings, we explored an alternative approach based on the primer exchange reaction (PER), an isothermal DNA polymerization reaction \cite{kishi_programmable_2018}. PER enables the addition of new DNA domains onto particle-grafted DNA using a DNA template in a hairpin structure as a catalyst \cite{moerman_simple_2023}. The template binds reversibly to the grafted DNA (primer), and in the presence of nucleotides and polymerase, its double-stranded region is copied, and the new domain is covalently attached onto the primer. The hairpin then detaches and can catalyze another reaction (Fig \ref{fig: PER}A).

The PER rate can be precisely controlled by adjusting the template concentration (SI \ref{SI: PER/reaction rate variation}) \cite{kishi_programmable_2018, moerman_simple_2023}, so we asked whether we could create DNA coating with tuneable composition by halting the reaction before it reaches completion. In the context of multispecific coatings, halting the initial reaction before completion leaves unreacted sites that can be targeted with a different DNA sequence in a subsequent functionalization step (Fig. \ref{fig: PER}B). Alternatively, multispecific coatings could be prepared through multiple simulataneous competing primer exchange reactions. Then the ratio of template concentrations would control the composition of the coating. 

\subsubsection{Multi-step PER}

\begin{figure*}[ht]
 \centering
 \includegraphics[width=17.1 cm,trim=0 330 0 0,clip]{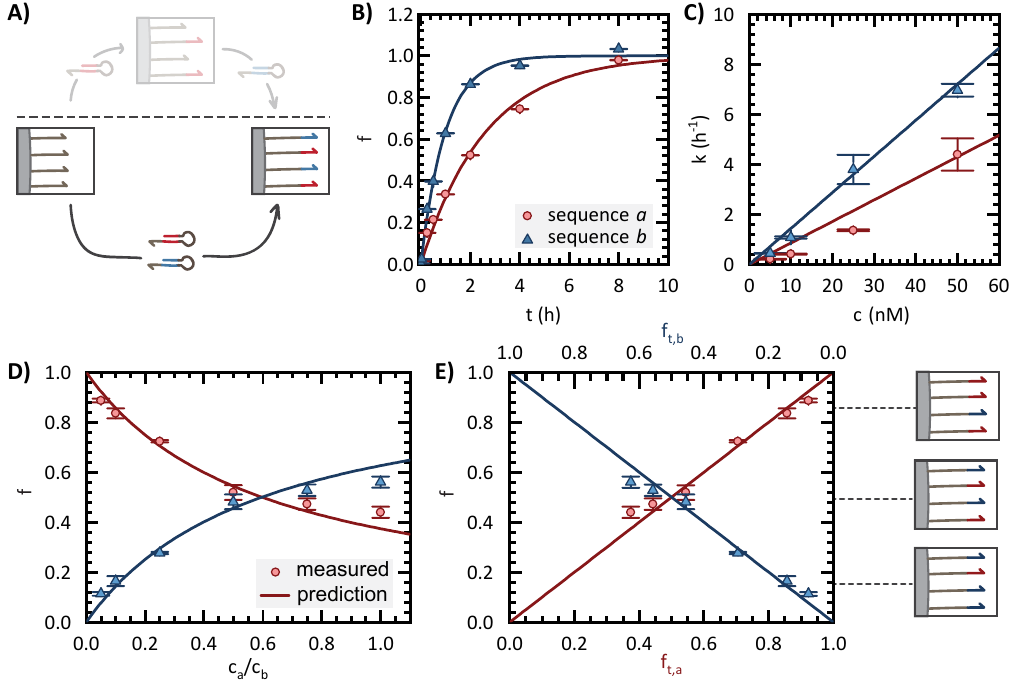}
 \caption{(A) Competitive PER using a multi-template design, where two different DNA domains are grown simultaneously on the same particle. (B) Grafting fractions of each sequence increase exponentially with reaction time, with sequence a reacting more slowly than sequence b. (C) Reaction rates scale linearly with template concentration, allowing control through template input. (D) In simultaneous PER, the grafting fraction of each sequence is tuned by the relative template concentrations. The x-axis represents the fraction $0 < c_a/c_b < 1.0$, corresponding to a fixed $c_a=10\,\mathrm{nM}$ and varying $0.5\,\mathrm{nM} < c_b < 10\,\mathrm{nM}$. 
For each particle, the grafting fractions $f_a$ and $f_b$ are measured independently, producing paired data points that are compared to the corresponding target compositions $f_{T,a}$ and $f_{T,b}$. 
Solid lines indicate predicted fractions based on rates from panel C, and the data points closely follow this trend. (E) Measured grafting fractions for both sequences agree well with their target ratios (solid lines), showing only minor deviations.}
 \label{fig: PER2}
\end{figure*}

Control over DNA coating composition in multi-step PER depends on precise control of completion in the initial step, which is governed by reaction time and rate. To target a specific completion in the first step, we measured the rates with which new DNA domains are grown on the particle as a function of the PER template concentration. To quantify these rates, we quenched PER at various time points, added fluorescent DNA complementary to the newly polymerized domains, and measured the fluorescent signal as a function of the reaction time in triplicate (SI \ref{SI: PER/FC data}, \ref{SI: PER/FC increase over time}). The results of these measurements are shown in Fig. \ref{fig: PER}C. Consistent with earlier work \cite{kishi_programmable_2018, moerman_simple_2023}, we found that the fraction of grafted DNA strands that contain the newly polymerized domains follows first order reaction kinetics and can be described by $f_a = 1-e^{-k_at}$, where $f_a$ represents the measured fraction of the DNA grafted onto the colloids onto which we grew a new functional domain $a$: $f_a=\rho_a/\rho_0$, with $t$ the reaction time, $\rho_0$ the total DNA grafting density measured from the plateau (SI \ref{SI: PER/FC increase over time}), and $k_a$ the template-concentration-dependent reaction rate. 

To assess the level of control that this method could give over the grafting density, we calculated an expected fractional conversion based on the reaction time $f_{T,a}$, the average reaction rate obtained from the fits, and the template concentration. We plotted the data from Fig. \ref{fig: PER}C on this rescaled x-axis in Fig. \ref{fig: PER}D. The plot shows a small deviation of each individual data point from the expected fractional conversion calculated from the average reaction rate, indicating that each reaction batch closely matched the conversion predicted from the model and suggesting that PER could give precise control over the composition with little batch-to-batch variation.

Next, to independently test to what extent the target grafting density can be obtained, we chose template concentrations and reaction times to target a series of specific grafting fractions: 0.23, 0.47, 0.65, 0.71, and 0.96, and plotted the results in Fig. \ref{fig: PER}D (purple traingles). Since the reaction was not completed in these experiments, the total number of DNA strands on the particle ($\rho_0$) cannot be obtained from the plateau. Instead we define the fractional conversion as $f_a = \rho_a/\rho_{ref}$, where $\rho_{ref}$ is the grafting density measured on a seperate reference sample that was fully extended to sequence $a$ under saturating PER conditions.

Fig. \ref{fig: PER}D shows that the independent data points closely match the target line, typically deviating by less than 10\%. These deviations are somewhat larger than those observed in the original datasets (red circles), which were used to determine the averaged $k_a$ values and target densities. This difference is likely due to variation in the time at which the reaction was stopped. Particularly at higher template concentrations and short reaction times, the reactions are faster and small variations in timing can substantially impact the final grafting fraction. Consistent with that notion the variation was larger for smaller target conversions.

Building on these results, we return to multispecific particles, which are generated by extending unconverted DNA strands with a second domain $b$ in a second reaction step. To test this approach, we grew a second domain $b$ onto the particles with $f_a=0.55$ from Fig. \ref{fig: PER}D, and measured the grafting density of both sequences using flow cytometry (SI \ref{SI: PER/multi-template and multi-step reaction}). After completion of the second step, domain $b$ was added to a fraction $f_b=0.31$, resulting in $f_{Total}=0.86$. Curiously, the total of the two sequences did not add up to 1. This mismatch could be due to variability in the final grafting densities observed between PER reactions, even when they are initiated from the same particle batch (SI \ref{SI: PER/FC increase over time}), or to the possibility that the grafting density of our reference particles is higher than that of the particles tested here ($\rho_0<\rho_{ref}$). Although this mismatch did not pose an issue for grafting a single sequence at a controlled density $\rho_a$, it limits control over the second sequence $\rho_b$, which was grafted at a substantially lower density than intended.

The uncertainty about the total grafting density poses a problem for making particles with many different sequences. As the number of unique grafted sequences increases, the number of required reaction steps increases accordingly, and the error is expected to accumulate. Each subsequent reaction step requires a reaction time that depends not only on the intrinsic reaction rate but also on the number of unreacted binding strands, which is in turn influenced by the previous grafting steps. Consequently, the spread in the coating composition is expected to grow as additional unique sequences are added.

Together, our results show that the sequential PER method provides good control over the grafting density of a single sequence, but is less suitable for multispecific DNA coatings. To sidestep issues associated with variations in $\rho_{total}$ and the propagation of errors in multiple PER steps, we next asked whether multispecific particles can be prepared by adding multiple sequences to a DNA coating simultaneously rather than in a stepwise manner.

\subsubsection{Competitive PER}

Adding multiple sequences to a DNA coating in a single step requires PER with multiple templates that compete for primers on the particles' surfaces (Fig. \ref{fig: PER2}A). In this approach, the composition of the DNA coating is set by the relative reaction rates with which the domains are grown onto the surface. The rates can be controlled with the relative concentrations of the templates, but also depend on the sequence of the DNA domain that is added~\cite{kishi_programmable_2018}.

To produce particles with well-defined ratios of two sequences in their DNA coatings, we first measured the inherent PER rate for both sequences, $k_a$, and $k_b$. Fig. \ref{fig: PER2}B, shows $f_a$ and $f_b$ as a function of reaction time at a template concentration of $c_a=c_b=10 $ $\nM$. It shows that sequence $b$ grows roughly twice as fast as sequence $a$, despite the fact that the binding domain is the same for both templates (gray in Fig. 3A). We found that this observation holds for a range of template concentrations (SI \ref{SI: PER/reaction rate all template concentrations}).

Such sequence dependence is consistent with earlier observations that growth rates depend strongly on the sequence of the domain that is polymerized \cite{kishi_programmable_2018, johnson_conformational_1993, murat_dna_2020}. These rate differences are therefore inherent to using multiple distinct sequences and cannot be eliminated by simple design choices. The ratio of template concentrations alone is therefore not sufficient to control the composition of the DNA coating; intrinsic sequence-dependent differences in polymerization rate must be corrected for. 

To correct for sequence-dependent differences in PER rate, we first measured the PER rate of both sequences as a function of the template concentration. Fig. \ref{fig: PER2}C shows that the PER rate increases linearly with template concentration for both sequences $a$ and $b$, but with different sequence-dependent slopes, $k_i'$: $k_i=k_i'c_i$. This finding suggests that, although different sequences grow at different absolute rates, their relative growth can still be controlled by adjusting template concentrations. To correct for the sequence dependence we assume that both primer exchange reactions proceed independently and express the fraction of the DNA on the colloid that turns into sequence $i$ as a function of both the inherent rates $k_i$ and $k_j$ and the template concentrations $c_i$ and $c_j$:

\begin{equation}
f_i = \frac{k_i' c_i}{k_i' c_i + k_j' c_j} \quad i=1,2 \quad \text{and} \quad i \neq j
\label{eq: f_i}
\end{equation}

Equation \ref{eq: f_i} gives the relative concentrations required to reach a target composition of the DNA coating $f_i$ and $f_j=1-f_i$. To test whether we can prepare multispecific particles with a target DNA composition, we mixed templates of sequences \textit{a} and \textit{b}, keeping the template concentration of sequence \textit{a} constant at $c_a = 10 \nM$ and varying the template concentration of sequence \textit{b} between $0.5~\nM$ and $10~\nM$. We compared the measured DNA composition with the prediction from Eq. 1 in Fig. \ref{fig: PER2}D and found a good agreement between the experimentally measured grafting fractions and the prediction from Equation 1, with deviations of up to 10\% from the target. This result shows that a sequence-dependent competitive disadvantage can be compensated for with increased template concentration.

The fact that rate-corrected template concentrations yield grafting ratios close to the targets shows that the two PER reactions behave effectively independently, despite competing for the same input strand. This independence suggests that template binding competition does not affect the relative reaction rates in the concentration range used here. We found that the simple proportionality breaks down, as expected, at high template concentrations, where the binding between template and primer becomes the rate limiting step (SI \ref{SI: PER/reaction rate all template concentrations}). Hence, we kept the sum of the template concentrations well below this limit ($\approx$ 50 nM).

To directly compare the mixed click approach with the competitive PER approach for producing particles with DNA coatings of specific compositions, we rescaled the x-axis of Fig. \ref{fig: PER2}D with Equation 1 to represent the intended grafting fraction (Fig. \ref{fig: PER2}E). This graph can be directly compared to Fig \ref{fig: DBCO}B, and shows that the spread around the targeted grafting fractions is consistently smaller for competitive PER than for mixed click. 

Taken together, we found that, with competitive PER, intended coating compositions can be produced consistently across a wide range of compositions with a precision of a few percent, but that doing so requires a systematic correction for the inherent rate differences between sequences through adjustment of the relative template concentrations during the reaction.

\subsection{Tuning Interaction Strength of DNA-Coated Colloids for Equilibrium Assembly}

Having established the precision with which the composition of DNA coatings can currently be controlled, we next ask what the consequences are for the self-assembly of multi-specific particles. How many independent DNA-mediated interactions that are reversible in the same temperature window can we reasonably engineer between one type of particle and other particles in the mixture? 

\begin{figure}
 \centering
 \includegraphics[width=8.3 cm,trim=0 355 0 0,clip]{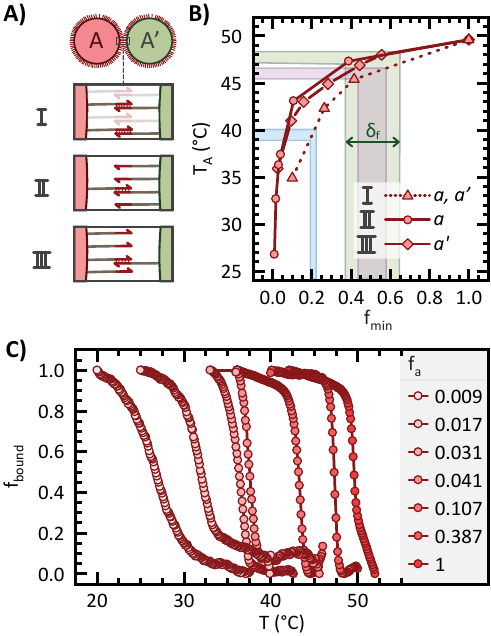}
 \caption{(A) Schematic representation of the binary system, where grafting densities were reduced either on (I) both binding partners equally or (II, III) on only one of the partners. (B) Measured assembly temperatures decrease steeply with lower grafting densities. For system I, this decline occurs at higher overall grafting densities, showing that the assembly temperature is determined not only by the partner with the lowest grafting density but by the combined grafting on both particles.  Shaded regions indicate the acceptable range of grafting densities yielding melting temperatures within 1 °C of the target value. The shading illustrates that this acceptable range narrows as the target fraction $f_a$ decreases (blue versus purple shading), and that the range is wider when the binding partner carries a higher overall grafting density (green versus purple shading). (C) Measured bound fraction of particles as samples were cooled from the singlet phase to the bound phase for different grafting fractions in system II (panel B). The assembly temperatures shown in panel B correspond to the temperatures at which 
$f_{bound}=0.5$.}
 \label{fig: Assemlby temperature}
\end{figure}

The first step to answering this question is to know how the melting temperature depends on the grafting density. Previous work showed that the melting temperature of DNA-coated colloids depends sub-linearly on grafting density when the grafting densities of both binding partners are equal \cite{wu_polygamous_2012, dreyfus_aggregation-disaggregation_2010, leunissen_switchable_2009} and that the temperature window in which the interactions are reversible is small, typically  1 °C \cite{wang_crystallization_2015, cui_comprehensive_2022, wu_polygamous_2012, jo_dna-coated_2024, oh_high-density_2015}. We found the same sublinear trend for particles with partial but equal coverage of DNA with sticky ends $a$ and $a'$, produced with the sequential PER method (triangles in Fig. \ref{fig: Assemlby temperature}b). 

In systems of multi-specific particles, the surface densities of sticky ends of two binding partners need not be equal though, which may affect the melting temperature associated with that DNA-mediated interaction. To study the effect of unequal grafting densities on the melting temperature, we varied only the coverage on the particles with sequence $a$ and measured the melting temperature of these particles mixed with the particles with the maximum grafting density of the sequence $a'$ (circles in Fig. Fig\ref{fig: Assemlby temperature}B), and vice-versa (diamonds in Fig\ref{fig: Assemlby temperature}B). Fig\ref{fig: Assemlby temperature}B shows the measured assembly temperature as a function of the lowest grafting density of both partners. Both measurements with unequal DNA grafting densities followed the same qualitative trend as those with equal grafting densities, and the temperature range over which the interaction remained reversible was similarly narrow (Fig. \ref{fig: Assemlby temperature}C). However, the melting temperatures increased more steeply at low grafting densities and transitioned earlier to a flattened, sub‑linear dependence, converging to the same melting temperatures. 

 How do we interpret this result? One might assume that the interaction strength between two DNA-coated colloids – and by extension the melting temperature – is determined by the minimum number of DNA connections that can form, and so by the lowest number of sticky ends between the two particles. Then the grafting density of the more densely coated particle should not matter, but Fig. 4B shows that it does. The difference between the curves with equal and unequal grafting densities in Fig. 4B is a consequence of the entropic contribution of binding between DNA-coated colloids; each strand on the lower density particle has multiple possible connection points on the other higher-density particle, which stabilizes the interaction\cite{dreyfus_aggregation-disaggregation_2010, leunissen_switchable_2009}.

The consequence of this effect for multi-specific particles is that the melting temperature is not only determined by the composition of the multi-specific particle in question but also that of its binding partner. This means that greater flexibility in DNA composition is available when targeting a specific melting temperature if the binding partner is fully coated with a single DNA type, rather than being multi‑specific.

Consider this numerical example to see the consequence of this effect. A particle with two types of DNA (each with $f_i=0.5$) in its coating has an acceptable range $\delta_f$ of $10\%$ in grafting density to still be within the 1 °C window around the target melting temperature when it’s binding partner has a full coating of the complementary DNA (green shading in Fig. 4B). When the binding partner also has two DNA types with each a fraction of 0.5, the acceptable range is only $5\%$ (purple shading). 

The constraints on coating composition become more stringent as more sequences are added to the coating because each additional sequence pushes the system closer to the steep regime of the melting curve. For example, if the particle in question has 5 different types of DNA in its coating with each a fraction of $f=0.2$, the acceptable variation in composition is only $0.5\%$ even when each binding partner carries only one type of DNA (blue shading). This constraint effectively limits practical implementation to two or three distinct sequences per particle.

We conclude that, although many orthogonal interactions could in principle be grafted on a single particle, for the purpose of equilibrium self-assembly we are much more limited; for DNA-coated colloids with grafting densities of $10^5$ strands per $\mu m^2$, and given the precision with which DNA coatings can be controlled, the number of reversible orthogonal interactions that can be engineered on one DNA-coated colloid is limited to two or three and only if those interactions are with mono-specific binding partners. This estimate assumes that each DNA strand interacts independently with no intra-particle interactions and that each sequence has the same free energy of association. The realizations of the equilibrium programmable, addressable assembly of many colloidal building blocks shown in simulations \cite{zeravcic_size_2014,mahynski_grand_2020,zeravcic_self-replicating_2014} thus require a substantial improvement in the maximum DNA density that can be obtained, the precision with which the coating composition can be controlled, or a widening of the 1 °C temperature window for equilibrium assembly, or fine-tuning methods to the melting temperature associated with each interaction through the addition of competitor or linker strands \cite{angioletti-uberti_re-entrant_2012, xia_linker-mediated_2020, dehne_reversible_2021, dehne_transient_2019, lowensohn_linker-mediated_2019}.

\begin{figure*}[ht]
 \centering
 \includegraphics[width=17.1 cm,trim=0 440 0 0,clip]{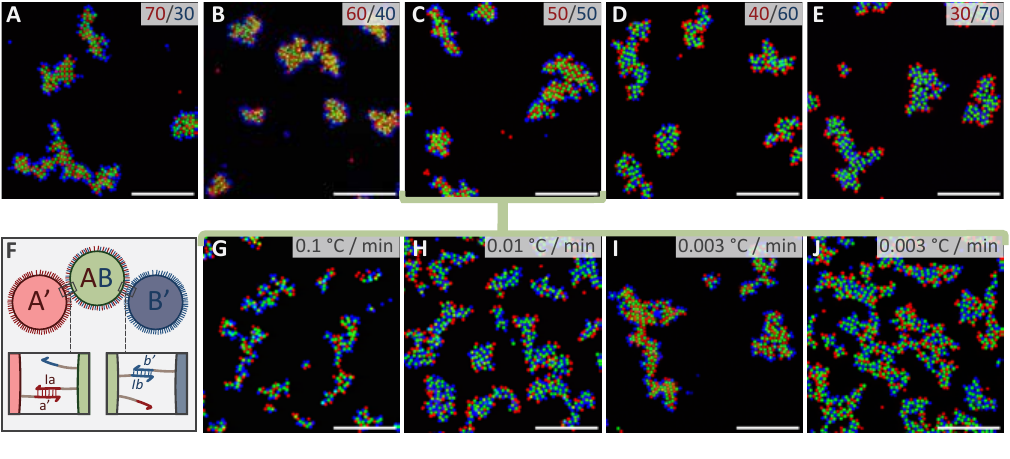}
 \caption{Various self-assembled structures of multispecific particles (green) with their complementary binding partners (red, blue). (A-E) Structures formed for different grafting fractions of the two sequences on the multispecific particle during a cooling ramp of 0.003 °C/min. Core–shell structures are observed, with the order of assembly switching depending on the grafting fractions. (F) Schematic illustrating the interactions between the respective particles. (G-J) Structures formed from 50/50 multispecific particles and their complements under different cooling ramps. Panels G-I show that the resulting structures depend on the cooling rate, with the order of assembly varying between rates. Panels I and J highlight that identical samples under the same cooling ramp can still yield different structures due to unavoidable sample and temperature variations.}
 \label{fig: Figure5}
\end{figure*}

\subsection{Equillibrium assembly of multispecific DNA-coated colloids}

Having shown that equilibrium assembly is practically limited to systems in which the relevant sequences can share a narrow assembly-temperature window, typically on the order of two or three DNA sequences on a particle, we now test explicitly whether our grafting-density control is sufficient to achieve such three-component assembly. To test this, we prepared a system of three particle types: one multi-specific, coated with sequences $a$ and $b$ and two mono-specific: one coated with only $a'$, and one with only $b'$. We varied the composition of the coating on the multi-specific particles and tested whether we could identify the regime in which all components co-assemble into a single, ordered structure. For the assembly experiments, we followed the typical annealing protocol required for crystallization \cite{wang_crystallization_2015, oh_high-density_2015, kim_engineering_2006, casey_driving_2012}: heat to above the melting temperature and cool slowly with a rate below 0.3 °C/hour. 

We found that after thermal annealing the particles assembled stepwise into core–shell structures (Fig. \ref{fig: Figure5}A-E). Such core-shell structures are expected when the binding energies between the multi-specific particles and their binding partners – and by extension the melting temperatures – are unequal\cite{di_michele_multistep_2013, wu_polygamous_2012}. In that case, the multi-specific particles first form dense clusters with one binding partner, and at lower temperatures, the second type of binding partner attaches to the previously formed clusters. Indeed the samples with $f_a=0.7$ and $f_a=0.6$ formed binary crystals in which the particles coated with $a$' are surrounded by particles coated with $b$' (Fig. \ref{fig: Figure5}A,B). The opposite core-shell structure formed when the multi-specific particles had the inverse composition $f_a=0.3$ and $f_a=0.4$. (Fig. \ref{fig: Figure5}D,E).

Particles with equal grafting density of both strands ($f_a=0.5$, $f_b=0.5$) still formed core-shell structures with the $a$' coated particles in the core (Fig. \ref{fig: Figure5}C). This asymmetry can be attributed to a small difference in free energy of hybridization between the two sequences (–13.98 kcal/mol for sequence \textit{a} vs. –12.94 kcal/mol for sequence \textit{b} for single strands free in solution, such that equal binding strengths are expected at a grafting ratio of 48/52 rather than 50/50 (SI \ref{SI: binding energies}). 

The difference between the target 50/50 DNA ratio and the composition of 48/52 that we predict to equalize the two binding strengths is smaller than the precision with which we can control the composition. Furthermore, this difference should result in a melting temperature difference well within the 1 °C assembly window indicated in Fig. \ref{fig: Assemlby temperature}, and yet we find that these particles did not co-crystallize under the applied thermal annealing protocol. Therefore, we asked next how the applied temperature protocol affected the assembly. We tested three different cooling rates (0.1°C/min, 0.01°C/min and 0.003°C/min) and observed markedly different structures for each. At relatively fast temperature ramps, mixed and disordered clusters formed (Fig. \ref{fig: Figure5}G), whereas slower ramps produced demixed structures in which crystalline domains of two particle types are surrounded by a shell of the third type. This contrast indicates a tradeoff in annealing rate that complicates equilibrium assembly of multi-specific particles: slower cooling promotes ordered structures, because it facilitates reversible binding and local reorganizations; but it also hinders co-assembly because it makes the system more sensitive to small differences in melting temperatures and the associated differences in nucleation rates (Fig. \ref{fig: Figure5}G,H). 

The comparison between the structures formed at various annealing rates also shows how sensitive the system is to such small differences in nucleation rate. Fig. \ref{fig: Figure5}H and Fig. \ref{fig: Figure5}I show the same sample cooled at 0.01°C/min and 0.003°C/min. Both produced core-shell structures but with opposite core- and shell-compositions. Fig. \ref{fig: Figure5}I and Fig. \ref{fig: Figure5}J show clusters formed at the same annealing rate (0.003°C/min) in a duplicate sample.  In one sample core-shell clusters formed while in the other the three particles co-assembled into a dense mixed structure with local crystalline domains. These results show how minute differences in particle concentration or in temperature (the heat conductivity of each sample cell might vary slightly) can affect the nucleation rates such that equilibrium co-assembly of even only three types of DNA-coated colloids requires substantial fine-tuning.

What do these findings mean for the experimental realization of simulations of fully addressable assembly? Even with precise control over the coating composition, minute deviations in local particle concentration, temperature, and interaction strength cause particles to tend to assemble in stages rather than simultaneously under a slow thermal annealing protocol. This effect is exacerbated by the collectivity of multivalent binding in systems of DNA-coated microparticles \cite{angioletti-uberti_understanding_2019, dreyfus_aggregation-disaggregation_2010}, but even in systems of DNA-coated nanoparticles and DNA origami building blocks the assembly rates strongly depend on temperature and binding strength \cite{sun_reversible_2005, stenke_growth_2022, zhan_recent_2023}. Together, that means that equilibrium calculations are insufficient for structure prediction in co-assembly of multiple building blocks and that assembly kinetics need to be considered \cite{jacobs_assembly_2025, di_michele_multistep_2013, furst_directed_2013}.

\section{Conclusions}

Realizing the full potential of DNA-coated particles for programmable self-assembly requires the functionalization of particles with multiple DNA sequences that interact orthogonally \cite{zeravcic_size_2014, mahynski_grand_2020,zeravcic_self-replicating_2014}. In this work, we showed that the primary limitation on the number of orthogonal equilibrium interactions that can be implemented is not the number of possible orthogonal DNA sequences that can be designed, nor the total number of unique DNA sequences that can be grafted. Instead, it is set by the requirement for equilibrium assembly that all interactions operate in the same narrow ($\approx$ 1 °C) temperature window in which DNA-mediated bonds are reversible.

Because the binding strength—and consequently the accessible temperature window—of each interaction depends on the surface density of each DNA strand type, variations in the DNA coating composition change the temperature at which each interaction is reversible. We compared a mixed click‑chemistry method with a PER‑based method for producing particles with mixed DNA coatings and found that both methods resulted in variations of approximately 10\% around the target density when the sequences in the coating had the same length. When the two sequences differed in length, the click‑based method could produce variations of more than 30\% from the target. Even in the best‑case scenario of a 10\% variation around the target composition, the steep dependence of DNA‑mediated interactions on grafting density means that designing particles with more than two types of equilibrium interactions encoded in their brush is not currently feasible.

Going beyond two equilibrium interactions requires strategies to fine-tune interaction strengths during self-assembly. Free-floating DNA strands could be used as linkers or competing strands \cite{angioletti-uberti_re-entrant_2012, xia_linker-mediated_2020, dehne_reversible_2021, dehne_transient_2019, lowensohn_linker-mediated_2019}, so that the concentrations of these strands would selectively modify one of the interaction strengths in the system to bring all transition temperatures in the same 1 °C window.
Removing the constraint of equilibrium interactions, the sequence space of DNA oligos of 11 nucleotides facilitates around 70 orthogonal interactions between colloidal particles \cite{wu_polygamous_2012}. This design space could be explored by facilitating local reorganizations using surface-mobile DNA linkers \cite{mcmullen_self-assembly_2022, rinaldin_colloid_2019} so that particles can reorganize locally even if the DNA-mediated interactions are not near the melting temperature. The property that multispecific particles will almost inevitably assemble sequentially—one interaction type at a time—upon thermal annealing could also be harnessed to drive folding‑based, line‑like colloidal assembly \cite{hecht_kinetically_2016} or core–shell assembly \cite{moerman_dna-encoded_2026}, without the need to modify the interactions on the go. Because the inherently pathway-dependent and non-equilibrium nature of such sequential multicomponent assembly precludes structure prediction from equilibrium considerations alone, simulation studies will be essential in establishing the design rules to exploit this behavior.

Beyond self-assembly, control over the composition of multi-specific DNA coatings is important for a range of applications. In biosensing multispecific coatings could enable the simultaneous detection of multiple targets \cite{buskermolen_continuous_2022,xu_whole-genome_2023,singh_dna-functionalized_2020}. More broadly, distinct DNA sequences on a single particle could perform distinct roles—such as localizing binding partners \cite{singh_dna-functionalized_2020}, initiating enzymatic signal production or amplification \cite{dehne_reversible_2021, dehne_transient_2019}, or triggering biological responses \cite{deng_atp-powered_2020,gines_microscopic_2017} —each with requirements for controlled grafting density.

\section{Experimental Section}

\subsection{Azide-functionalization of polystyrene particles}

The azide‑functionalized PS‑PEO polymer was synthesized following a modified version of the protocol by Oh et al. \cite{oh_high-density_2020}, where a detailed procedure is provided. Briefly, PS‑PEO (Polymer Source; PS 3800 g/mol, PEO 6500 g/mol) was first sulfonylated by reacting it with methanesulfonyl chloride in anhydrous dichloromethane and triethylamine under ice‑bath conditions, followed by stirring overnight at room temperature. The polymer was then precipitated and washed three times in diethyl ether/methanol to remove residual reagents. In the second step, the sulfonyl‑modified PS‑PEO was reacted with sodium azide in anhydrous dimethylformamide at 65 °C for 24 h. The resulting azide‑modified polymer was again precipitated and washed in diethyl ether/methanol, after which the supernatant was removed and the purified PS‑PEO‑N\textsubscript{3} was freeze‑dried and stored at –20 °C until use.

Next, polystyrene particles were coated with PS‑b‑PEO‑N\textsubscript{3} as follows. The PS‑b‑PEO‑N\textsubscript{3} polymer was dissolved in Milli‑Q water at 4 mg/mL and sonicated overnight. In a glass vial, 150 µL of polystyrene particles (2.6 wt\%, 1.1 µm; PolySciences Polybead Sulfate Microspheres, nominal diameter 1.00 µm) was mixed with 86 µL of the polymer solution. The vials were placed on an orbital shaker for 1 h, after which 125 µL of tetrahydrofuran (THF) was added. The samples were shaken for 1 h with the lids closed, then the lids were removed and shaking was continued for an additional hour. Subsequently, 1 mL of Milli‑Q water was added, the vials were closed again, and the samples were left on the shaker overnight. The particles were then washed by centrifugation, the supernatant was replaced with Milli‑Q water, and the particles were stored at 4 °C.
For confocal microscopy (Fig. \ref{fig: Figure5}), particles were fluorescently labeled with Bodipy 493, Bodipy 633, or Rhodamine B to distinguish different species. The dyes were prepared at 2 wt\% in toluene and 5 µL was of dye solution was added during the polymer‑functionalization protocol simultaneously with the THF addition.

\subsection{DNA Coupling to Particles Using Click Chemistry}

Polystyrene particles were functionalized with single-stranded DNA via strain‑promoted azide–alkyne click chemistry. A suspension containing 0.1 wt\% particles and 3.5 µM DBCO-functionalized DNA (Integrated DNA Technologies; custom HPLC-purified oligonucleotides supplied at 100 µM in TE buffer consisting of 10 mM Tris and 0.1 mM EDTA), supplemented with 1 M NaCl and 0.05 wt\% Pluronic F127, was heated to 65 °C and continuously mixed for 24 h to promote covalent attachment of the DBCO-DNA to the azide-functionalized particles. After the reaction, the particles were washed five times with MilliQ water and stored at 4~°C.

Particles were  prepared either with a single DNA sequence, by adding 3.5 µM of one type of DBCO-DNA, or with a mixture of two different sequences at varying ratios while keeping the total DNA concentration at 3.5 µM. The single‑sequence functionalization was used to prepare complementary partners in multispecific assemblies and to generate the input particles for the primer exchange reaction experiments. The two‑sequence functionalization was used to produce multispecific DNA coatings. An overview of all sequences used is provided in SI \ref{SI: DNA sequences}.

\subsection{Attachment of new DNA groups using primer exchange reactions (PER)}

PER reactions were performed on DNA‑coated particles to grow new DNA domains as follows. A PER reaction premix was first prepared at 2.5× the target concentration, containing 2.5× ThermoPol DNA polymerase buffer (New England Biolabs, supplied at 10× concentration), 31.25 mM MgCl$_2$ (New England Biolabs; 100 mM stock solution), and 250 µM each of dATP, dCTP, and dTTP (ThermoFisher, 10 mM stock solutions).

To measure PER reaction rates, a 5 µL reaction mixture was prepared consisting of 0.5 µL of 1 wt\% DNA-coated particles (final concentration: 0.1 wt\%), 2 µL of the premixed PER buffer, 0.5 µL of template solution (concentrations varied between 5 and 100 nM depending on the experiment), and MilliQ water to reach a total volume of 4 µL. The reaction was initiated by adding 1 µL of 0.8 U/µL Bst Large Fragment DNA polymerase (New England Biolabs), yielding a final enzyme concentration of 0.16 U/µL. Samples were incubated at 20~°C for 8 hours and aliquots were taken at multiple time points. At each time point, a 0.5 µL aliquot was removed and quenched in 0.1 M EDTA to halt enzymatic activity. Particles were then washed three times with Milli‑Q water and stored at 4 °C.

For the preparation of multispecific particles using multi‑step PER, reactions were halted by adding an excess of 0.1 M EDTA solution at defined time points, which were varied to target different DNA grafting densities. For the preparation of multispecific particles using competitive PER, the same protocol was followed except that two different hairpin solutions were added in varying ratios. The hairpin solutions were premixed before adding the particles to ensure homogeneous DNA binding. Reactions proceeded for 8 h, after which the samples were diluted with 45 µL of 0.1 M EDTA, washed with Milli‑Q water, and stored at 4 °C.

\subsection{Flow Cytometry Experiments}

Flow cytometry was used to quantify DNA grafting densities. Samples were prepared by mixing 5 µL of 0.01 wt\% particles with 5 µL of 1 µM fluorescently labeled DNA complementary to the grafted DNA sequence in 40 µL of 1 M NaCl in TE buffer. The mixture was incubated at 4~°C for at least one hour, diluted in 200 µL of the same buffer, vortexed and analyzed using a BD FACS Symphony A3 flow cytometer. Only signals from single particles were collected by applying a gating strategy based on forward and side scattering. Fluorescence was measured for 10,000 particles.

For multispecific particles, complementary fluorescent DNA was added separately for each sequence, and fluorescence was measured independently for each label. To determine the grafting fraction, the fluorescence intensity of multispecific particles was compared to that of two sets of monospecific particles functionalized with the maximum grafting of the corresponding sequences.

\subsection{Microscopy Experiments}

Glass capillaries (CM Scientific 0.1 x 2.00 mm) were cleaned by sonicating in isopropanol for 10 min, dried under nitrogen flow, and treated with plasma for 10 min. The capillaries were then placed vertically in a glass jar containing 200 µL of 1,1,1,3,3,3-hexamethyldisilazane (HMDS) and left to silanize until the solution fully evaporated (approximately 1 day). Afterward, they were stored in acetone. Before use, capillaries were washed with ethanol and dried under nitrogen flow.

For the measurements shown in  Fig. \ref{fig: Assemlby temperature}, samples were prepared at 0.1 wt\% of each particle type in TE buffer containing 250 mM NaCl and 1 wt\% Pluronic F127. Samples were loaded into the capillaries, the ends were sealed with grease, and the capillaries were attached to a coverslip (epredia 24 x 60 mm, \# 1) using UV-curable glue. A temperature ramp of 0.1 °C/min was applied using a Linkam heating stage, and samples were imaged simultaneously in bright field using a Olympus BX53M microscope . Assembly temperatures and singlet fractions were extracted from changes in the average image intensity associated with particle cluster formation. Normalized image intensities (0–1) quantified the bound particle fraction $f_{bound}$. The assembly temperature ($f_{bound}=0.5$) was obtained from a sigmoidal fit to the intensity–temperature data.

For the measurements shown in Fig. \ref{fig: Figure5}, samples were prepared similarly but with 0.1 wt\% multispecific particles and 0.05 wt\% of each complementary particle in TE buffer containing 100 mM NaCl and 3 wt\% Pluronic F127. Temperature ramps were applied either continuously at 0.1 °C/min or stepwise, using increments of 0.3 °C or 0.1 °C every 30 min. all ramps were run from 48 °C to 38 °C. The final structures were imaged using a Leica SP8 confocal microscope.

\section*{Author contributions}
T.C.M.S. and P.G.M. conceived the experiments. T.C.M.S. and A.S. performed the experiments and analysis. T.C.M.S. and P.G.M. wrote the manuscript. All authors discussed the results and provided feedback on the experiments and manuscript. 
\\

\section*{Acknowledgements}
We thank Barbara Malheiros for assistance with the preparation of particles used for DNA functionalization. We also thank Barbara Malheiros, Bahar Rouhvand, and Dominique Leemans for useful discussions. P.G.M acknowledges funding from NWO through VI.Veni.222.194.

\section*{References}   

\bibliographystyle{apsrev4-2} 
\bibliography{references} 

\clearpage

\setcounter{section}{0}
\setcounter{subsection}{0}
\setcounter{figure}{0}
\setcounter{table}{0}
\setcounter{equation}{0}

\begin{widetext}

\section*{\textbf{Supplementary Information}}
\section{DNA sequences} \label{SI: DNA sequences}

\begin{table}[h!]
\centering
\begin{tabular}{lll}
\hline
\textbf{Name} & \textbf{Sequence (5'-3')} & \textbf{Figures} \\
\hline
a & /5DBCON/ 40T ACCATCCTACC & 1 \\
polyT & /5DBCON/ 40T & 1 \\
b (short) & /5DBCON/ 40T ACTCATCTCAA & 1 \\
b (long) & /5DBCON/ 40T C TCCTAATTC ACTCATCTCAA & 1 \\
I & /5DBCON/ 40T ACTTCACTT & 2,3,4 \\
template a & ACCATCCTACC GGGCCTTTTGGCCC GGTAGGATGGT AAGTGAAGT /3InvdT/ & 2,3 \\
template b & ACTCATCTCAA GGGCCTTTTGGCCC TTGAGATGAGT GAATTAGGA /3InvdT/ & 3\\
Ia & /5DBCON/ 40T ACTTCACTT ACCATCCTACC & 2,3,4,5 \\
Ib & /5DBCON/ 40T ACTTCACTT ACTCATCTCAA & 3 \\
Ia' & /5DBCON/ 40T ACTTCACTT GGATGGT & 4,5 \\
a' & /5DBCON/ 40T GGATGGT & 5 \\
b' & /5DBCON/ 40T TTGAGATG & 5 \\
imager a & GGT AGG ATG GTA A/36-FAM/ & \\
imager b & /56-FAM/TT GAG ATG AGT & \\
\hline
\end{tabular}
\caption{Oligonucleotide sequences used in the study. “40T” indicates a stretch of 40 thymidines. The table also indicates which sequences were used in each figure.}
\label{tab:oligos}
\end{table}

\clearpage

\section{DNA grafting density of monospecific colloids prepared by DBCO-click chemistry} \label{SI: DBCO/seq dependent grafting density}

We measured the maximum grafting density $\rho_{total}$ for several sequences using a titration experiment. Particles coated via DBCO-click chemistry (under saturation conditions to achieve maximum coating density) were mixed and left to hybridize with fluorescently labeled complementary DNA strands. Samples were prepared with increasing amount of fluorescent DNA. The fluorescent intensity was measured by flow cytometry and averaged over 10.000 particles. The results are shown in Fig. \ref{fig: FigureSI1}. We observed a linear increase of the fluorescent intensity until a plateau was reached. The transition to a plateau indicates saturation of fluorescent DNA binding to the particle. The data were fitted with 

\begin{equation}
    F = \frac{q}{2K c_p} \left( \left(c_i K + c_p N K +1 \right) = \sqrt{\left( c_i + c_p N K +1 \right)^2 - 4K^2 c_i c_p N} \right)
    \label{eq: titration curve}
\end{equation}

which follows from equillibrium binding. Here $c_i$ is the imager concentration, $c_f$ the particle concentraiton, $K$ is the sequence depedent hybridization constant and $q$ and $N$ are fitting parameters corresponding to the fluophore efficiency and the number of DNA strands per particle, respectively. 

\begin{figure}[ht]
    \centering
    \includegraphics[width=17.1cm,trim=0 470 0 0,clip]{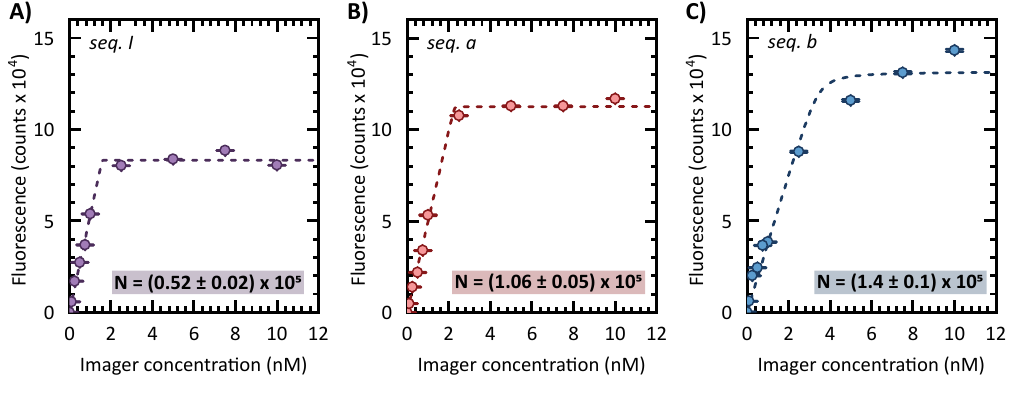}
    \caption{The plots show the increase in fluorescence intensity as a function of the concentration of fluorescent (imager) DNA added to the suspension for (A) sequence $I$, (B) sequence $a$ and (C) sequence $b$. The data were fitted using Equation \ref{eq: titration curve}, and the number of DNA strands per particle, $N$, was determined from the onset of the plateau.}
    \label{fig: FigureSI1}
\end{figure}

\clearpage

\section{Flow cytometry data (fluorescence histograms) of multispecific particles prepared by DBCO click chemistry} \label{SI: DBCO/FC data}

\begin{figure}[ht]
    \centering
    \includegraphics[width=0.6\linewidth,,trim=0 100 0 0,clip]{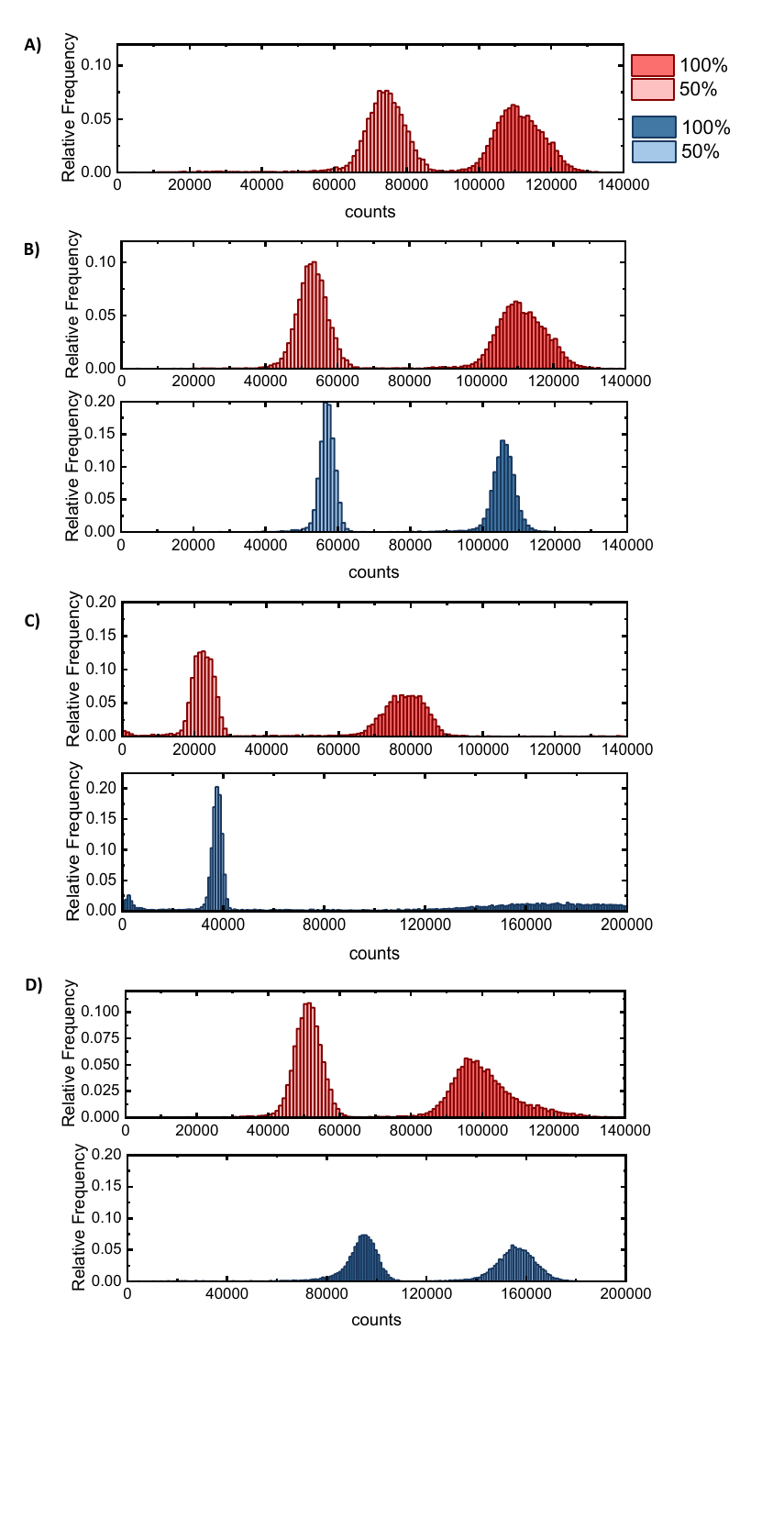}
    \caption{Representative fluorescence histograms for four sequence combinations targeting a grafting fraction of 0.5, shown alongside reference particles with a grafting fraction of 1. Fluorescense histograms are given for both seq \textit{a} (red) and the second sequence that is varied (blue); (A) seq. a + poly-T (not measured), (B) seq. a + seq. b (short), (C) seq. a + seq. b (long), (D) seq. a + seq. c (short, no poly-T spacer). Histograms reflect fluorescence from 10,000 individual particles measured after hybridization with fluorescent complementary strands. Differences in histogram width and intensity arise from variations in fluorescent labels and hybridization energies; notable cases include the broad reference distribution in (C) and enhanced fluorescence in (D), likely due to local sequence environment. Because the grafting of the two sequences is coupled, interpretation focuses on the red-channel data, which used a constant imager and exhibited Gaussian-like distributions, indicating that deviations from target grafting fractions are not attributable to particle-to-particle heterogeneity.}
    \label{fig: DBCO histogram multispecific}
\end{figure}

\section{Fluorescence measurements of primer-exchange-reactions}

\subsection{Flow cytometry histograms during a primer-exchange reactions} \label{SI: PER/FC data}

To quantify the rate of new DNA domain growth on particle-grafted DNA (Fig. \ref{fig: PER2}C), we took aliquots at different time points during a primer-exchange reaction. The particles were then hybridized with an excess of complementary fluorescently labeled DNA, and the increase in fluorescence was measured over time for 10,000 particles. An example for a primer-exchange reaction using sequence $a$ at a template concentration of 10 nM (Fig. \ref{fig: PER}C, \ref{fig: SI/PER fluorescence}) is shown in Fig. \ref{fig: SI figure -- PER FC }.

\begin{figure}[ht]
    \centering
    \includegraphics[width=0.35\linewidth,trim=0 0 0 0,clip]{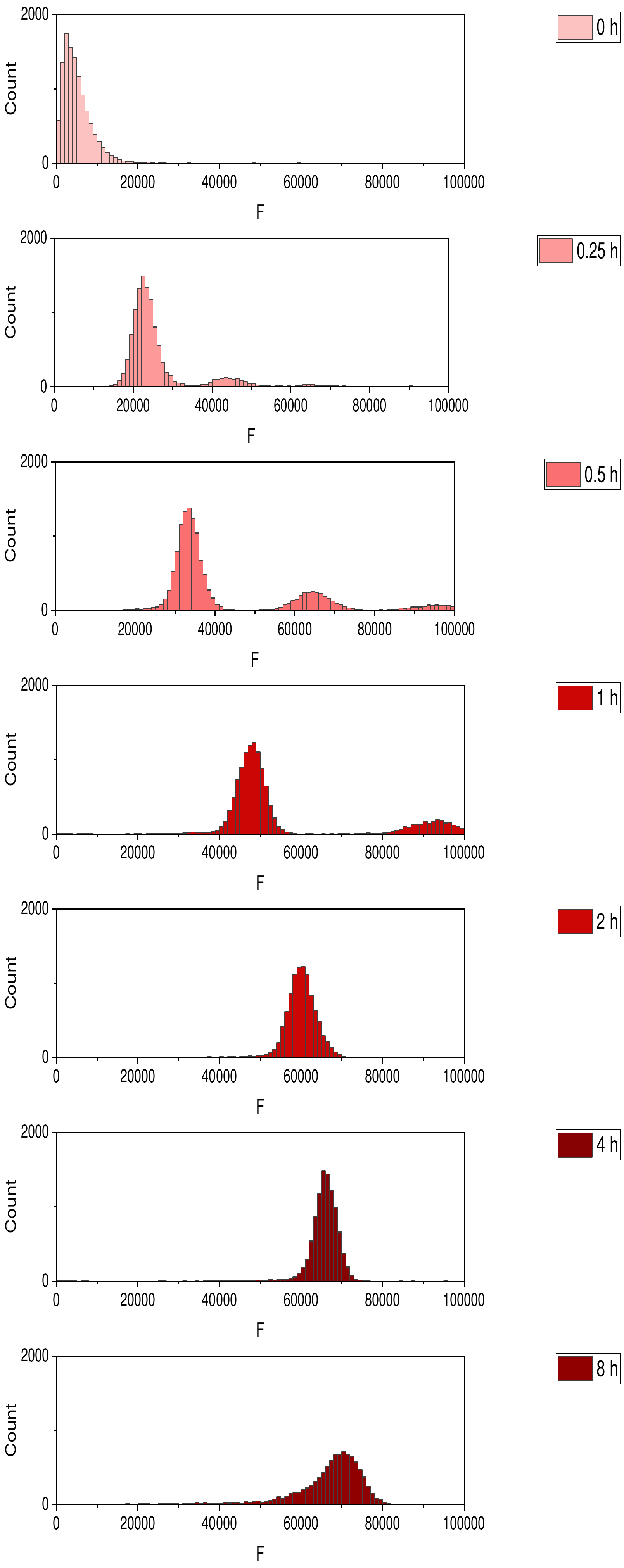}
    \caption{Histograms show the distribution of fluorescence measured for 10,000 particles hybridized with complementary fluorescent DNA during flow cytometry at different reaction times. Typically, two Gaussian peaks are observed: a main peak and a smaller, higher-fluorescence peak that can be attributed to larger particles formed by aggregation during sample preparation. From the main peak, we extract the average fluorescence for each condition and monitor the increase over time.}
    \label{fig: SI figure -- PER FC }
\end{figure}

\clearpage

\subsection{Fluorescence increase over reaction time for primer-exchange reaction for different template concentrations} \label{SI: PER/FC increase over time}

Fig. \ref{fig: SI/PER fluorescence} shows the averaged fluorescence signal measured over 10,000 particles, plotted as a function of reaction time. These values were obtained from the fluorescence histograms measured by flow cytometry (Fig. \ref{fig: SI figure -- PER FC }). Here, the fluorescence reflects the number of fluorescent DNA sequences bound to the particles, which in turn corresponds to the number of DNA groups extended with the new sequence \textit{a}. Regardless of template concentration, the fluorescence increase can be described by \[F = F_{sat}(1-exp(-kt))\]. This equation was fitted to the experimental data to obtain the template-concentration-dependent rate $k$ and the saturation fluorescence $F_{sat}$, which represents the maximum grafting density on a particle. Interestingly, although all reactions were started from the same batch of particles, we observed that $F_{sat}$ varies slightly, indicating some variation in the final grafting densities. The grafting fractions shown in Fig. \ref{fig: PER}C are defined relative to the maximum grafting for each batch: \[f_a = {F}/{F_{sat}}\]. 

\begin{figure}[ht]
\centering
    \includegraphics[width=17.1cm,trim=0 440 0 0,clip]{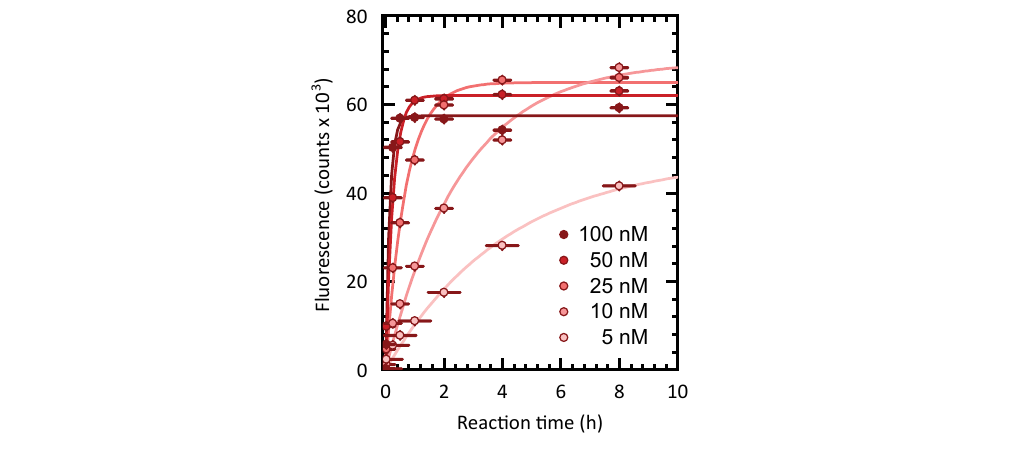}
    \caption{Fluoresent signal measured over 10000 particles with flow cytometry plotted against the reaction time for several template concentrations and fitted to obtain values for the rate $R$ and saturation fluorescence $F_{sat}$. }
    \label{fig: SI/PER fluorescence}
\end{figure}

 \clearpage

\section{Primer-exchange reaction rates in triplicate} \label{SI: PER/reaction rate variation}

\begin{figure}[ht]
    \centering
    \includegraphics[width=\linewidth,trim=0 480 0 0,clip]{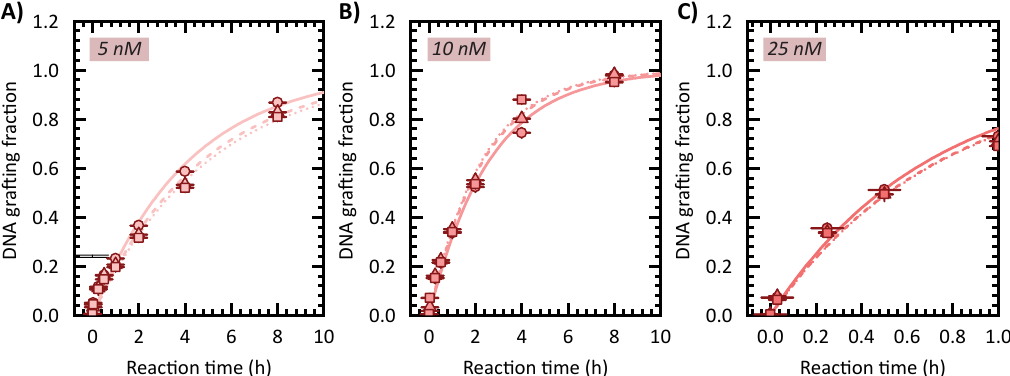}
    \caption{Reaction rates of the primer-exchange reaction measured in triplicate for template concentrations of (A) 5 nM, (B) 10 nM and (C) 25 nM, all for sequence \textit{a}. Small variations in rate were observed under identical conditions, likely arising from differences in enzyme activity. These variations introduce deviations of a few percent in the achievable grafting fraction and are most relevant when reactions are not run to completion. In multi-template primer-exchange reactions, both templates are expected to be affected equally. }
    \label{fig:enter-label}
\end{figure}

\clearpage

\section{Primer-exchange reaction kinetics}

\subsection{Template concentration dependent reaction rates for sequences \textit{a} and \textit{b}} \label{SI: PER/reaction rate all template concentrations}

Reaction rates for the primer-exchange reactions were measured for both sequence \textit{a} and sequence \textit{b} at template concentrations ranging from 5 to 100 nM. For both sequences, the reaction rate increases approximately linearly with template concentration, consistent with the trend discussed in Fig. \ref{fig: PER2}C. We initially observe faster polymerization rates for sequence \textit{b} (blue) than for sequence \textit{a}. This difference can be attributed to sequence dependence of the polymerization reaction \cite{kishi_programmable_2018, johnson_conformational_1993, murat_dna_2020} or to differences in template strand release and rebinding kinetics.
Interestingly, at higher template concentrations, no further increase in the reaction rate is observed for sequence \textit{b}. We interpret this as the regime in which template release and rebinding are no longer rate-limiting, and the polymerization step itself becomes the rate-limiting process. The fact that this regime appears for sequence \textit{b}, but not for sequence \textit{a} within the tested concentration range, is consistent with the hypothesis that the faster rates of sequence \textit{b} arise from more rapid template turnover. In that case, saturation of the turnover process would occur at lower template concentrations.
In the context of preparing multispecific particles via PER, we advise avoiding high template concentrations. In multi-step reactions, high reaction rates amplify deviations from the target composition due to limited time precision. In competitive reactions, the plateau in the rate makes it difficult to fine-tune the relative reaction rates of different sequences. 

\begin{figure}[ht]
    \centering
    \includegraphics[width=\linewidth,trim=0 480 0 0,clip]{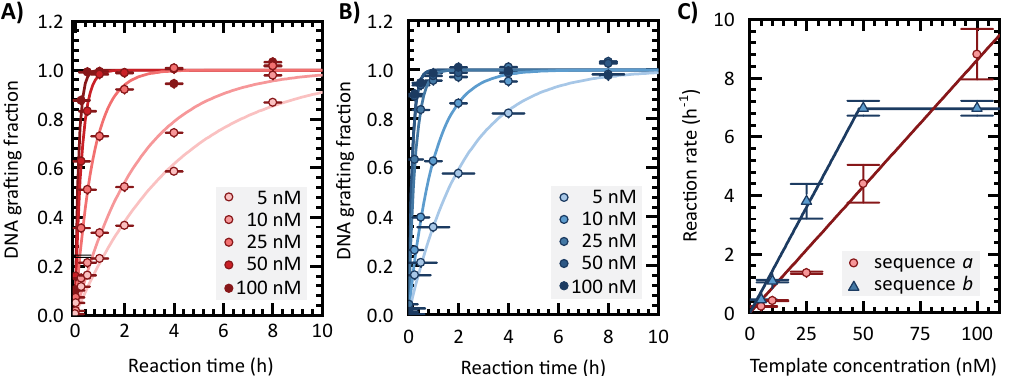}
    \caption{ (A) Reaction rates measured for all template concentrations of sequence \textit{a}. (B) Reaction rates measured for all template concentrations of sequence \textit{b}. (C) Reaction rates calculated from the fits of panels A and C. Reaction rates increase linearly with template concentration, as shown by the blue data at some point the reaction rate flattens to a maximum, we hypothesize that this is because at high concentration the reaction is no longer limited by the template concentration but by the time required to grow new DNA domains. To control the composition of multispecific coatings it's essential to stay away from these high template concentrations.}
    \label{fig: SI PER2}
\end{figure}

\clearpage

\begin{figure}[ht]
    \centering
    \includegraphics[width=17.1cm,trim=0 500 20 0,clip]{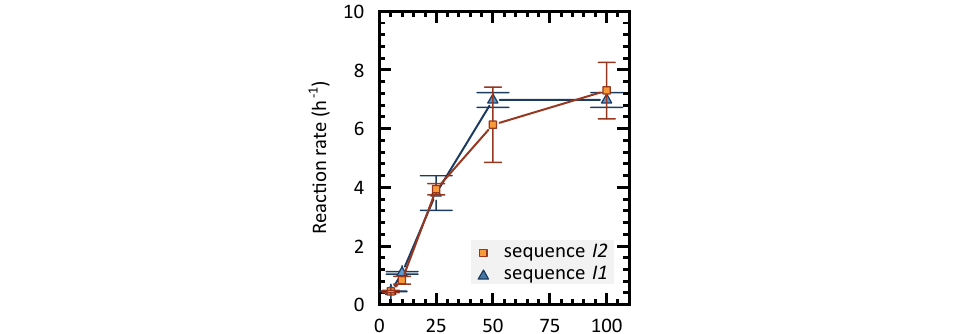}
    \caption{Rates were measured for two different starting domains (I1: ACTTCACTT and I2: CTCCTAATTC), where in both cases sequence $b$ was grown. The measured rates show that the growth rate is independent of the exact starting domain, at least when the initial sequences have similar length and binding energy.}
    \label{fig: SI PER3}
\end{figure}

\clearpage

\subsection{Master curve upon rescaling}

Because the reaction rate $k$ scales linearly with template concentration (Fig. \ref{fig: SI PER2}C), the conversion curves obtained at different template concentrations (Fig. \ref{fig: SI PER2}) can be collapsed onto a single master curve by rescaling time with the measured rate constant $kt$. Specifically, plotting the grafting fraction as a function of the dimensionless time causes the data from all template concentrations to overlap (Fig. \ref{fig: SI PER mastercurve}). 
This collapse indicates that the primer-exchange reaction follows the same underlying kinetic mechanism across the explored concentration range. The template concentration primarily sets the timescale of the reaction, without altering the functional form of the growth dynamics. 

\begin{figure}[ht]
    \centering
    \includegraphics[width=17.1cm,trim=0 500 0 0,clip]{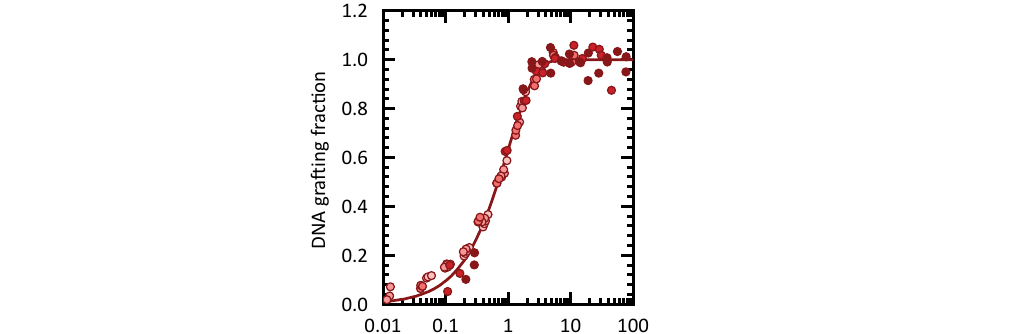}
    \caption{Rescaled primer-exchange reaction kinetics. Grafting fraction as a function of rescaled time $kt$ for different template concentrations. The rate constant $k$ was obtained from exponential fits to the fluorescence increase. Data collapse onto a single master curve upon rescaling.}
    \label{fig: SI PER mastercurve}
\end{figure}

\clearpage

\section{Multi-step and multi-template reactions} \label{SI: PER/multi-template and multi-step reaction}

Fig. \ref{fig: SI multi-step and competitive FC data} shows the increase of grafting fraction of sequences \textit{a} and \textit{b} during a multi-step PER and competitive PER reaction. Both approaches were designed to target grafting fractions of $f_a=f_b=0.5$. In the multi-step reaction (Fig. \ref{fig: SI multi-step and competitive FC data}A), sequence \textit{a} slightly exceeded the target, reaching $f_a=0.52$. After initiation of the second step, sequence \textit{b} reached a final grafting fraction of $f_b=0.31$, below the intended value. In the simultaneous multi-template reaction (Fig. \ref{fig: SI multi-step and competitive FC data}B), both sequences increased at comparable rates but again reached unequal final grafting fractions, with $f_a=0.42$ and $f_b=0.30$.
The deviation from the target values is likely related to a lower overall maximum grafting density of the particle batches used for the multispecific coatings compared to the reference particles used to calibrate the reaction conditions.

\begin{figure}[ht]
    \centering
    \includegraphics[width=0.72\linewidth,trim=0 420 120 0,clip]{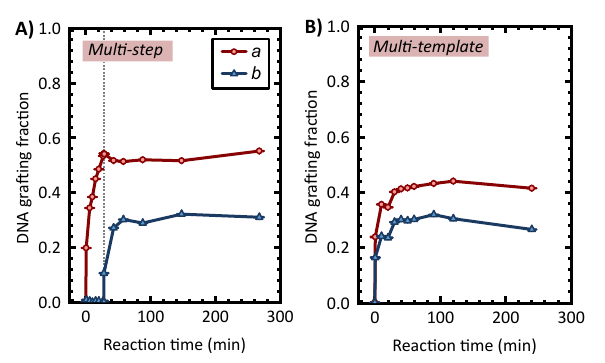}
    \caption{(A) Grafting fractions of sequences \textit{a} and \textit{b} during a multi-step PER reaction. Sequence \textit{a} is grown first; after 28 min (dashed line), the reaction is stopped and a second reaction is initiated to grow sequence \textit{b} on the remaining sites. (B) Grafting fractions of sequences \textit{a} and \textit{b} during a multi-template PER reaction in which both sequences are grown simultaneously. }
    \label{fig: SI multi-step and competitive FC data}
\end{figure}

\clearpage

\section{Binding energy calculations} \label{SI: binding energies}

 To calculate the free energy of hybridization of strands in free in solution using NUPACK \cite{zadeh_nupack_2011} -- using T = 20 C, a salt concentration of 1M NaCl and DNA concentration of 1 \uM . For the sequence combination $Ia$ and $a'$ this gave a binding free energy of –13.98 kcal/mol, and for the sequence combination $Ib$ and $b'$ this gave a binding free energy of –12.94 kcal/mol. Assuming that DNA strands bind independently, in the absence of significant steric or cooperative effects, the total interaction free energy can be approximated as $\Delta G_{total} = n_{bonds} \Delta G_{hyb}$. Under this approximation, we calculate that, to equalize the total binding energies, the grafting fractions must satisfy $f_a = \frac{13.98}{12.94+13.98};=0.52$ and $f_b = \frac{12.94}{12.94+13.98};=0.48$. 
 ]
 
\clearpage

\section{Additional microscopy images for figure \ref{fig: Figure5}}

\subsection{Control experiments of multispecific particles with both binding partners} \label{SI: Assembly/crystals with seperate partners}

\begin{figure}[ht]
    \centering
    \includegraphics[width=17.1cm,trim=0 500 0 0,clip]{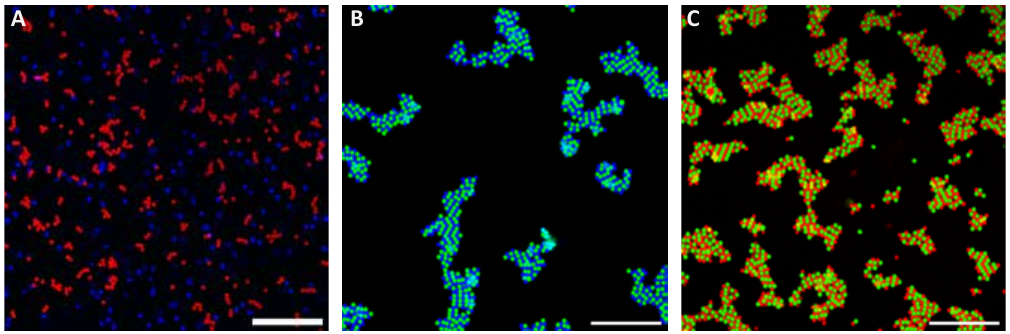}
    \caption{(A) Complementary particles coated with sequences a' (red) and b' (blue) mixed in the absence of the multispecific binding partner do not interact. Occasional small clusters of red particles are observed and attributed to weak, non‑specific DNA hybridization. (B,C) Multispecific particles (blue) coated with grafting fractions of $f_a=0.53$ and $f_b=0.47$ for both sequences a and b were mixed separately with binding partners coated with the complementary sequences a’ (red) and b’ (blue), respectively. Both samples were then cooled slowly, in steps of 0.3 °C per hour, to induce crystallization. The microscopy images show the resulting structures after this temperature ramp, revealing crystalline order. These results indicate that multispecific particles retain their ability to crystallize with either binding partner.}
    \label{fig: SI figure -- crystals }
\end{figure}

\clearpage

\subsection{Self-assembly structures of additional grafting densities}

\begin{figure}[ht]
    \centering
    \includegraphics[width=17.1cm,trim=0 250 0 0,clip]{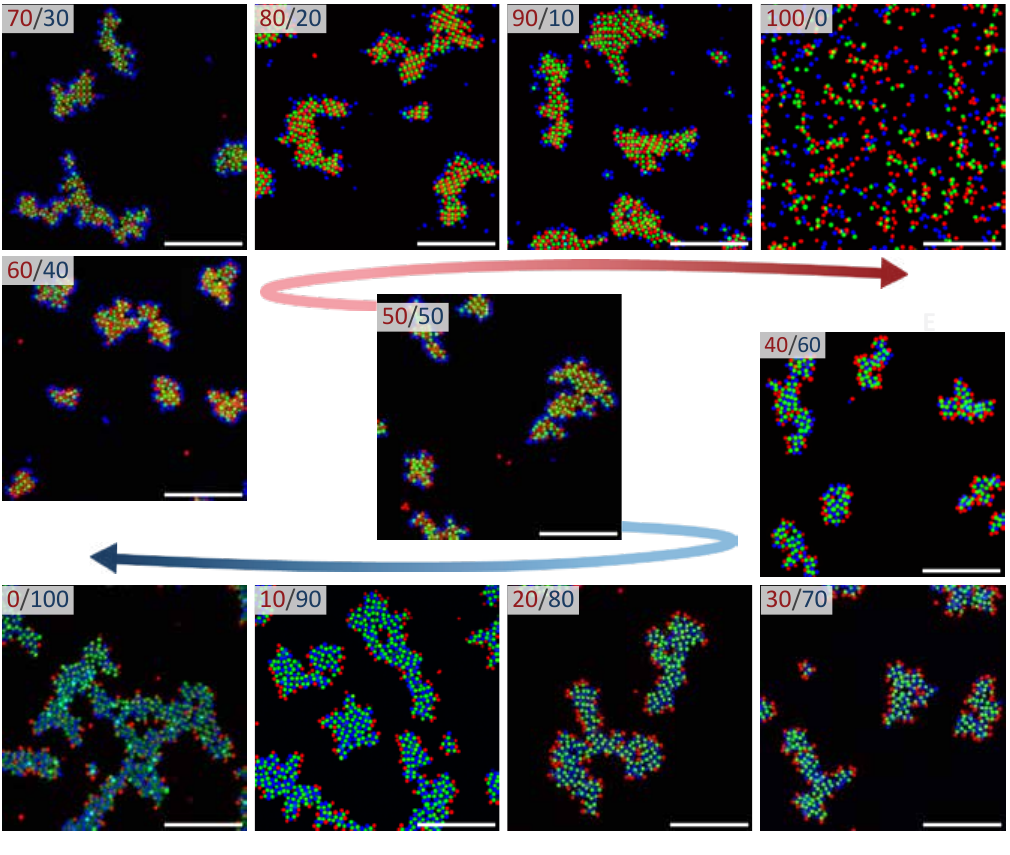}
    \caption{Various self‑assembled structures formed by multispecific particles (green) mixed with their complementary binding partners (red, blue). The structures arise from samples cooled at 0.003 °C/min and prepared with different grafting fractions of the two sequences on the multispecific particles. Core–shell architectures are observed, with the order of assembly switching depending on the grafting fractions. At higher grafting densities, crystalline domains appear within the inner cluster.}
    \label{fig: SI figure -- full ramp }
\end{figure}

\clearpage

\subsection{Self-assembly structures under faster cooling rates}

\begin{figure}[ht]
    \centering
    \includegraphics[width=15.5cm,trim=0 0 0 0,clip]{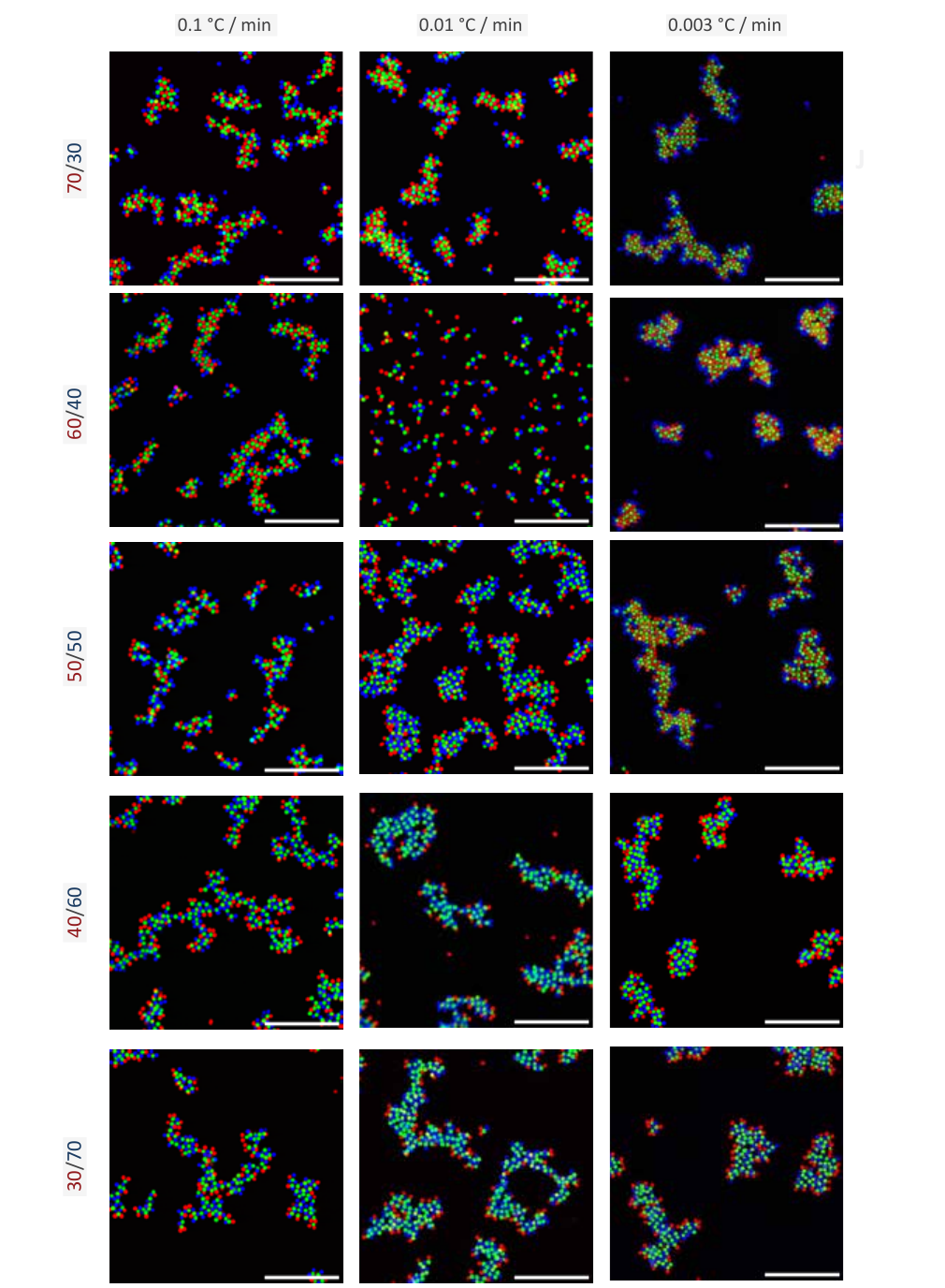}
    \caption{Structures formed from multispecific particles (green) and their complementary partners (red, blue) under different cooling ramps and for various DNA grafting ratios. Faster cooling rates produce more disordered, mixed structures, with the switch in assembly order occurring gradually. Slower cooling ramps yield more ordered, compositionally separated clusters, with a sharper transition in assembly order as the grafting ratio is varied.}
    \label{fig: SI figure -- all ramps }
\end{figure}

\end{widetext}

\end{document}